\documentclass[12pt,final]{iopart}
\usepackage{iopams}
\usepackage{graphicx}
\usepackage[breaklinks=true,colorlinks=true,linkcolor=blue,urlcolor=blue,citecolor=blue]{hyperref}

\begin{document}

\title{Comment on `An elementary argument for the magnetic
field outside a solenoid'}

\author{Ben Yu-Kuang Hu}
\address{Department of Physics, The University of Akron, Akron, OH 44325-4001, U.S.A.}
\ead{byhu@uakron.edu}

\begin{abstract}
An alternative elementary argument for the magnetic field outside a solenoid is described.

\end{abstract}

\vspace{2pc}

\submitto{European Journal of Physics}
%\maketitle

\section{Introduction}

Pathak\cite{pathak17} showed that the superposition of magnetic fields produced by the individual coils of a solenoid tend to cancel outside the solenoid.  The author stated that ``Informally, the two ends of the long solenoid behave like two magnetic poles, each of which produce that fall off with the inverse square of distance, and thus the fields outside are small when the length of the solenoid is increased, keeping the current fixed."  In this comment, I point out that this argument can be made rigourous and formal.
%, and can be used to provide analytic results for the magnetic field both inside and outside the solenoid.

The magnetic field that is produced by time-independent electric currents is given by the Biot-Savart law
\begin{equation}
\mathbf B(\mathbf r) = \frac{\mu_0}{4\pi} \int dV'\ \frac{\mathbf J(\mathbf r')\times \mathbf (\mathbf r - \mathbf r')}{\vert\mathbf r - \mathbf r'\vert^3},
\end{equation}
where $\mathbf J$ is the charge current density from arises from both free and bound currents.
Bound currents arise from the magnetisation $\mathbf M$ of materials and are given by
$\mathbf J_b = \nabla\times \mathbf M$.  In the case of the boundary of a magnetised material, the $\mathbf J_b$ becomes surface current of $\mathbf K_b = \mathbf M \times \hat{\mathbf n}$ where $\hat{\mathbf n}$ is the unit vector pointing out of and perpendicular to the magnetised surface boundary.\cite{Griffiths}

The magnetic field $\mathbf B$ produced by a given charge current density does not depend on whether the currents are free or bound.  Thus, the magnetic field due to a given free current configuration $\mathbf J_f(\mathbf r)$ can be obtained as follows.  First, one obtains the magnetisation $\mathbf M(\mathbf r)$ which gives bound currents that reproduce $\mathbf J_f$ ({\em i.e.}, a $\mathbf M(\mathbf r)$ such that $\nabla\times \mathbf M =\mathbf J_f$).  Then, $\mathbf B$ be found from $\mathbf M$ using techniques borrowed from electrostatics, as described below.

\section{Obtaining the magnetic field from the magnetisation}

In the absence of free currents and in static situations, the Maxwell's equations for the auxiliary magnetic field $\mathbf H$, which is defined as
\begin{equation}
\mathbf B \equiv \mu_0(\mathbf H + \mathbf M) \label{eq:defineH}
\end{equation}
are
\begin{eqnarray}
\nabla \times \mathbf H &=& 0\label{eq:3}\\
\nabla \cdot \mathbf H &=& -\nabla\cdot\mathbf M.\label{eq:4}
\end{eqnarray}
These equations mirror the Maxwell's equations for electrostatics in the absence of free charge,
\begin{eqnarray}
\nabla \times \mathbf E &=& 0\label{eq:5}\\
\nabla \cdot \mathbf E &=& -\frac1{\epsilon_0} \nabla\cdot\mathbf P,\label{eq:6}
\end{eqnarray}
where $\mathbf E$ is the electric field and $\mathbf P$ is the electric polarisation.  Eq.~(\ref{eq:6}) comes from $\nabla\cdot\mathbf D = 0$ (in the absence of free charge) and $\mathbf D = \epsilon_0\mathbf E + \mathbf P$.

The $-\nabla\cdot\mathbf P$ inside the volume of an electrically polarised material represents the bound charge density $\rho_b$.  At the the surface of the polarised object, the polarisation is discontinuous, and $-\nabla\cdot\mathbf P$ reduces to $\mathbf P\cdot \hat{\mathbf n}$ where $\hat{\mathbf n}$ is the unit vector pointing perpedicular to and out of the surface.\cite{Griffiths3}  Therefore, the electric field due to the electric polarisation is given by Coulomb's law (expressed in terms of the electrostatic potential $\varphi$)
\begin{eqnarray}
\mathbf E(\mathbf r) &=& -\nabla \varphi(\mathbf r)\label{eq:7}\\
\varphi(\mathbf r) &=& \frac1{4\pi\epsilon_0} \int_{\mathcal V} dV'\ \frac{\rho_{b}(\mathbf r')}{\vert \mathbf r- \mathbf r'\vert} +
\frac1{4\pi\epsilon_0} \int_{\mathcal S} dS'\ \frac{\sigma_{b}(\mathbf r')}{\vert \mathbf r - \mathbf r'\vert}\label{eq:8}\\
\rho_b(\mathbf r) &=& -\nabla\cdot\mathbf P(\mathbf r)\label{eq:9}\\
\sigma_b(\mathbf r) &=& \mathbf P(\mathbf r) \cdot \hat{\mathbf n}\label{eq:10}
\end{eqnarray}
In Eq.~(\ref{eq:8}), $\mathcal V$ and $\mathcal S$ are the volume and the surface of the region containing the polarised material.

Since Eqs.~(\ref{eq:3}) and (\ref{eq:4}) can be obtained from Eqs.~(\ref{eq:5}) and (\ref{eq:6}) by replacing $\mathbf E \rightarrow \mathbf H$ and $\epsilon_0^{-1}\mathbf P \rightarrow \mathbf M$, it follows that $\mathbf H$ can be obtained from $\mathbf M$ by the same replacement in Eqs. (\ref{eq:7}) -- (\ref{eq:10}).\cite{Helmholtz}  This together with Eq.~(\ref{eq:defineH}) gives
\begin{eqnarray}
\mathbf B(\mathbf r) &=& \mu_0 \Bigl(-\nabla \phi^*(\mathbf r) + \mathbf M (\mathbf r)\Bigr),\label{eq:11}\\
\phi^*(\mathbf r) &=& \frac1{4\pi} \int_{\mathcal V} dV'\ \frac{\rho^*_{b}(\mathbf r')}{\vert \mathbf r- \mathbf r'\vert} +
\frac1{4\pi} \int_{\mathcal S} dS'\ \frac{\sigma^*_{b}(\mathbf r')}{\vert \mathbf r - \mathbf r'\vert}\\
\rho^*_b(\mathbf r) &=& -\nabla\cdot\mathbf M(\mathbf r)\\
\sigma^*_b(\mathbf r) &=& \mathbf M(\mathbf r) \cdot \hat{\mathbf n}.
\end{eqnarray}
These relationships have been derived in alternative ways in Refs.~\cite{RMC} and \cite{zangwill}.

%Thus, the magnetic field for the case of permanently magnetised material can be obtained by simply using the techniques for obtaining the electric field in the presence of polarisation to get $\mathbf H$, and then adding $M$ and multiplying by $\mu_0$ to get $\mathbf B$.

\section{Magnetic field of a finite uniform solenoid}

Let us apply the results obtained above to the case examined in Ref.~\cite{pathak17}, that of a finite-length solenoid of uniform surface current density.
Assume that one has an idealised solenoid of arbitrarily shaped cross section with the axis in the $z$-direction a surface current $\mathbf K$ of uniform magnitude.  (For a physical solenoid where the current is carried by a wire that is repeatedly wound uniformly around the solenoid, $\vert\mathbf K\vert = nI$, where $I$ is the current through the wire and $n$ is the number of loops of wire per unit length.)

\begin{figure}[htb!]
 \centering
 \includegraphics{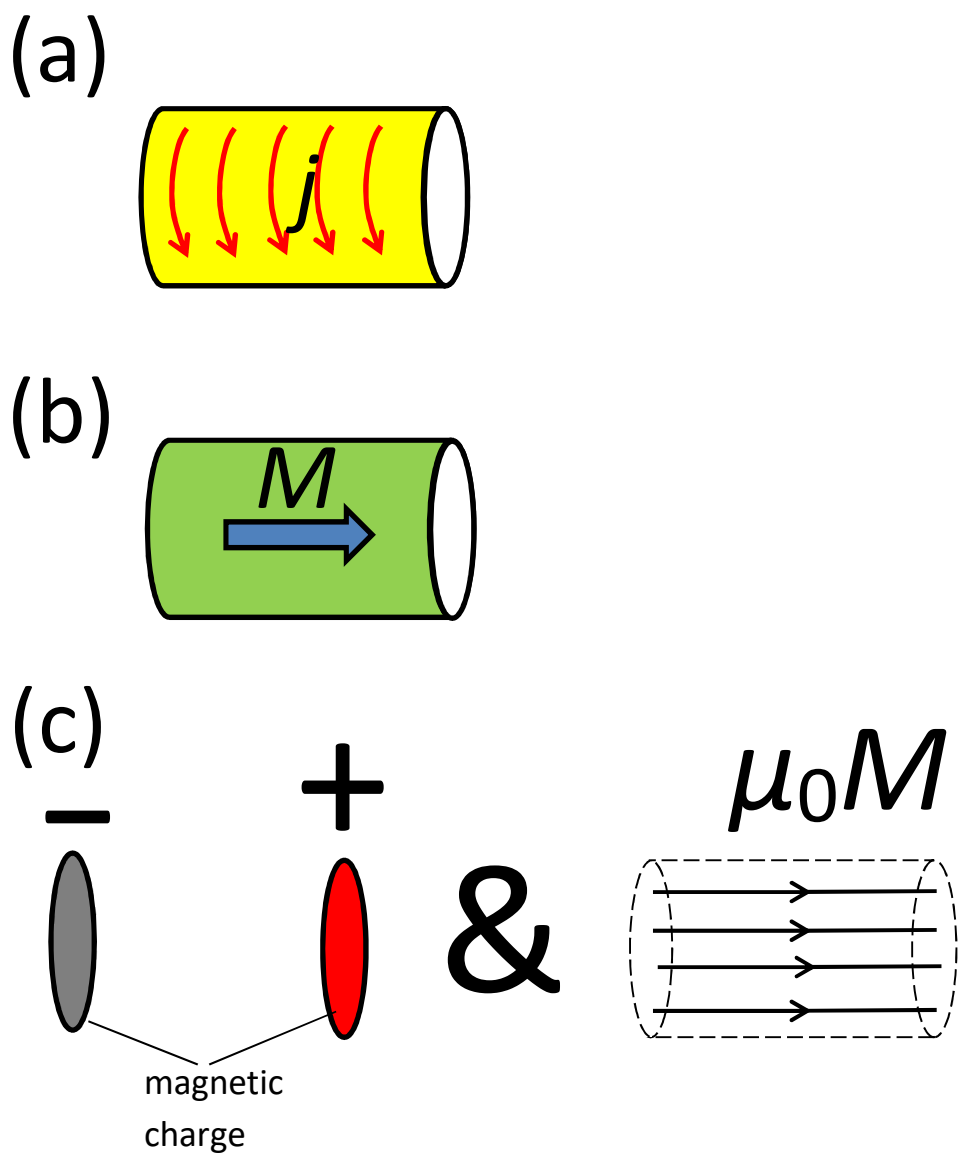}
 \caption{Procedure for reducing the problem of obtaining the magnetic field of a solenoid to an electrostatic problem.  (a) The original problem: Solenoid with uniform current.  (b) Magnetisation which reproduces current shown in part (a).   (c) Magnetic field obtained from the magnetisation shown in part (b).}\label{figureone}
\end{figure}

The surface current for the solenoid is reproduced by a permanent magnetised material of constant magnetisation $\mathbf M = K\hat{\mathbf z}$ that is the same shape as the solenoid.  Since the magnetisation is constant, the bulk magnetic charge $\rho^*_b = -\nabla\cdot\mathbf M = 0$,
The surface magnetic charge for this configuration is a disc of uniform density  $\sigma_b^* = -\vert \mathbf M\vert$ on the south pole of the magnet and $\sigma^* = \vert \mathbf M\vert$ on the north pole.   This shows rigourously that the magnetic field outside the solenoid is identical in form to that of the electric field of two discs of uniform charge density $\pm \sigma$ at the ends of the solenoid, with $\sigma/\epsilon$ replaced by $\mu_0 \sigma_b^* = nI\mu_0$.  As indicated by Eq.~(\ref{eq:11}), inside the solenoid, one needs to add $\mu_0\mathbf M$, which in this case is $\mu_0 K \hat{\mathbf z}$. This procedure for obtaining the magnetic field is shown schematically in Fig.~\ref{figureone}.

For a finite solenoid where the length scale of the cross section is much smaller than other length scales, the magnetic charges at the ends of the solenoids can be approximated by point charges.  Using Coulomb's law, this gives Eq.~(5) in Ref.~\cite{pathak17} for the magnetic field outside the solenoid.   In the case of a solenoid with a circular cross section, a semi-analytic expression for the magnetic field can be obtained as an expansion in spherical harmonics, using the solution of the electric field of a uniform charged disc.\cite{Griffiths1}  This expression was obtained by Muniz {\em et al.}\cite{muniz15} using a different technique.

\section{Another example}

The method described here assumes that one can come up with $\mathbf M$ which satisfies $\nabla\times\mathbf M = \mathbf J_f$, which is not necessarily easy to do.   In cases where there are only surface currents, as in solenoids, it is typically easier to guess the appropriate $\mathbf M$ which reproduces the currents.   Here is another example, which is related to a paper by Ref. \cite{ferreira18} on the magnetic field produced by an axial current along a finite segment of a cylindrical tube.

\begin{figure}[htb!]
 \centering
 \includegraphics{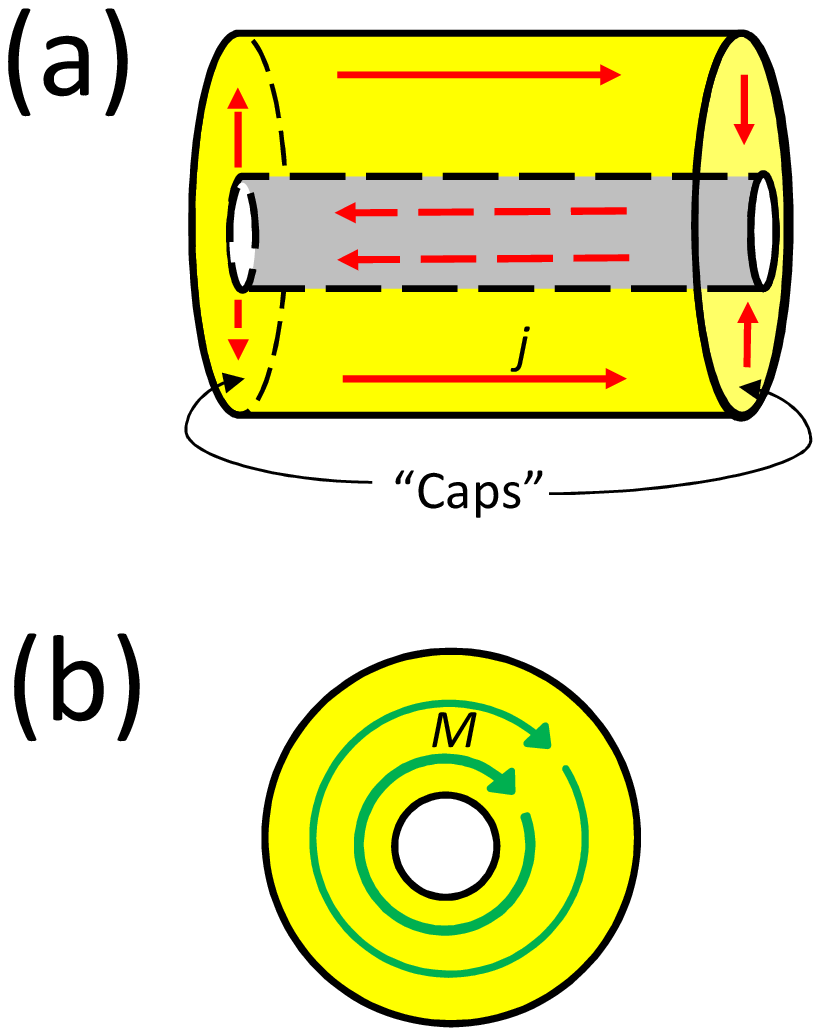}
 \caption{ Magnetic field for current flowing on surfaces of co-axial cylinders.  (a)  The current flows in the axial direction, and flows along the ``caps" at the ends between the inner and the outer cylinders.  (b) Magnetisation which reproduces current shown in part (a), as viewed along the axes of the cylinders.}\label{figuretwo}
\end{figure}

%In Ref. \cite{ferreira18}, the magnetic field was found to be non-zero everywhere in space (while of course decreasing in magnitude with increasing distance from the cylinder).
Let us consider two coaxial cylindrical conductors that are capped with conducting materials at both ends, so that the two cylinders are electrically connected, as shown in Fig.~\ref{figuretwo}.  A uniform current flows in the axial direction on the outside cylinder, radially inwards along the cap to the inner cylinder, in the opposite axial direction on the inner cylinder, and then radially outwards back to the outer cylinder at the other cap.

By charge conservation, the magnitude of the surface currents must be inversely proportional to $s$, the distance from the axis of the cylinders.  Let the total current flowing through these cylinders be $I_t$.  Then, the surface current density is $I_t/(2\pi s)$.

In this case, the currents are reproduced by the magnetisation $\mathbf M = I_t/(2\pi s)\; \hat{\boldsymbol\phi}$ (where $\hat{\mathbf \phi}$ is the unit vector in the azimuthal direction) in between the cylinders and zero outside.  Since $\nabla\cdot\mathbf M = 0$ and $\mathbf M\cdot\hat{\mathbf n} = 0$, there is no magnetic charge.  Hence, the magnetic field is $\mathbf B = \mu_0 \mathbf M = \mu_0 I_t/(2\pi s) \hat{\boldsymbol\phi}$ in between the cylinders and zero everywhere outside.

Ref.~\cite{ferreira18} showed that axial currents in the cylindrical segments alone ({\em i.e.}, excluding the caps) produce non-zero magnetic fields everywhere in space.  Therefore, everywhere outside the cylinders, the radial currents on the caps connecting the two cylinders produce magnetic fields that exactly cancel the magnetic field due to the axial currents along the cylinders.
This result might appear at first sight to be surprising -- why does this exact cancelation of magnetic fields occur outside the cylinders?  This unexpected result can be understood by recognising that this example is in fact a toroid with a rectangular cross section (the axis of symmetry of the toroid is the co-axes of the cylinders).  As is well-know, the magnetic field of any uniform toroid of arbitrary cross section is confined within the toroid itself.\cite{Griffiths2}

In fact, the method described in this comment can used as an alternative way to to prove this fact.  Consider a toroid with an arbitrary cross section and a uniform surface current.  In cylindrical coordinates, the interior of the toroid is given by a certain region in the $s$--$z$ plane.  The current on the surface of the toroid is reproduced by the magnetization $\mathbf M = (I_t/2\pi s) \hat{\boldsymbol \phi}$ inside this region and zero outside (here $I_t$ is the total current around the toroid; {\em i.e.}, if there is a current $I$ in a wire that is wound $N$ times around the toroid, then $I_t = N I$).  Since $\nabla\cdot \mathbf M = 0$ in the interior and $\mathbf M \cdot\hat{\mathbf n} = 0$ on the surface of the toroid, there is no magnetic charge contribution to the magnetic field, and therefore $\mathbf B = \mu_0 \mathbf M = (\mu_0 I_t/2\pi s)\ \hat{\boldsymbol \phi}$ inside of the toroid and zero outside.

\end{document}